\documentclass[aps,prl]{revtex4}
\def\ee{\end{equation}}
\def\bea{\begin{eqnarray}}

\begin{document}

\title{Environmental Collapse Models}
\author{Adrian \surname{Kent}}
\affiliation{Centre for Quantum Information and Foundations, DAMTP, Centre for
  Mathematical Sciences, University of Cambridge, Wilberforce Road,
  Cambridge, CB3 0WA, U.K.}
\affiliation{Perimeter Institute for Theoretical Physics, 31 Caroline Street North, Waterloo, ON N2L 2Y5, Canada.}
\email{A.P.A.Kent@damtp.cam.ac.uk} 

\date{June 6, 2022} 

\begin{abstract}
  We propose dynamical collapse models in which the stochastic collapse
  terms affect only photons and/or gravitons.   In principle, isolated
  systems comprising only massive particles could evolve unitarily indefinitely
  in such models.   In practice, since photons and gravitons are
  ubiquitous and scatter from massive particles, dynamical collapses
  of the former will effectively induce collapses of the latter.

  In non-relativistic models in which particle number is conserved and
  interactions are modelled by classical potentials, massive systems
  can be modelled as collections of elementary massive particles bound
  by potentials, interacting with an environment of photons and
  gravitons.  In this picture, although the photon and/or graviton collapse
  dynamics effectively localize massive systems, these collapses
  take the effective form of approximate measurements on the
  environment whose effect on the massive systems is indirect.
  We argue that these environmental collapse models, like
  standard mass-dependent spontaneous localisation models, may
  be consistent with quantum experiments on
  microscopic systems while predicting very rapid effective collapse of
  macroscopic massive systems, and hence a potential solution
  to the quantum measurement problem.
  However, the models considered here have different experimental
  signatures from standard mass-dependent spontaneous localisation
  models.  For example, they predict no deviations from standard
  quantum interferometry for mesoscopic systems of massive particles isolated
  from a decohering environment, nor do they predict anomalous  
  spontaneous emission of radiation from isolated matter of the
  type prediction by standard mass-dependent spontaneous localization models.
  New experiments and analyses are
  required to obtain empirical bounds on the decoherence rate in
  our models.     

\end{abstract}
\maketitle
  
\section{Introduction}
 
In previous papers \cite{kent2012real,kent2015lorentzian,kent2017quantum}, we
described ideas aimed towards defining realist and
Lorentz covariant versions of relativistic quantum
theory or quantum field theory.   
A key idea in these papers is to postulate a fundamental split
$H = H_A \otimes H_B$ in the Hilbert space of nature, together with
a suitable asymptotic late time behaviour of the
degrees of freedom in one factor.
Together with initial conditions on the full state, these are used 
to define a realist ontology for quantum theory.  
For example, Refs. \cite{kent2015lorentzian,kent2017quantum} propose
an ontology for quantum  matter defined by the initial state together with
the outcomes of fictitious measurements on the electromagnetic
or gravitational field at asymptotically late times.
In simple models this ontology
aligns with common intuitions and with the language commonly used
to describe experiments.   For example \cite{kent2017quantum}, the ontology
suggests that, if an unstable excited atom can radiate a photon to
infinity without scattering or absorption, it reaches its ground state at
an almost precisely defined time that is Poisson distributed,
with mean given by the expected photon emission time.

Here we use related intuitions to propose a different type of
model.
Roughly speaking, we take $H_A$ to be the degrees of freedom for all massive
particles and $H_B$ for massless particles.
Alternatively, we could take $H_B$ to correspond to force carrier particles.
However, we here consider simple non-relativistic models and are primarily
interested in the particles associated with long-range forces.
We treat hadrons as elementary and ignore gluons; we also ignore W, Z and
Higgs bosons.
We thus take $H_A$ to be everything but photons and/or gravitons and
$H_B$ to be electromagnetic and/or gravitational degrees
of freedom.   We consider dynamical collapse models in which the
stochastic ``collapse terms'' act on $H_B$ but not $H_A$.
Because matter and radiation are generally entangled, these collapses
nonetheless affect matter indirectly.

For example, a $10^{-5}$m diameter dust grain 
interacts with a flux of $\sim 10^{13}$ thermal photons per second 
at $300$K.   If the dust grain is in a superposition of
two macroscopically distinct position states, for which
a collapse model predicts that collapses swiftly occur in a way that
distinguishes the corresponding states of scattered radiation,
then the grain will be swiftly localized to one position state. 
As this illustrates, the effective spontaneous collapse
rate implied for matter depends on its environment, its scattering signature, and the parameters of a photon
(or graviton) collapse model.

\section{Non-relativistic photon collapse}

Ref. \cite{kent2017quantum} is not a dynamical collapse model, but
does consider asymptotic late time fictitious measurements of photons
that could be interpreted as the electromagnetic field undergoing
spontaneous collapses, albeit only at an asymptotically late time. 
Dynamical photon collapses were subsequently
considered by Pearle \cite{pearle2018dynamical}
within the framework of non-relativistic mass-dependent continuous spontaneous localization
(CSL) models.   Mass-dependent non-relativistic CSL models are motivated by the
observation that mass density is a natural preferred observable.   This choice
avoids the need to distinguish elementary from composite particles
and allows parameter choices that appear consistent with experimental
data to date.   Pearle's motivation for considering photon collapse in this
context was that, if consistent relativistic versions of these models
exist, it seems natural that they would relate collapse to stress-energy and
plausible that the relevant quantity is the total stress-energy, to which
particles of zero rest mass also contribute. 
Pearle thus suggests that a non-relativistic limit might
relate collapse to energy density (for both photons and
massive particles) rather than mass density (which of course
is relevant only for massive particles).
To implement this, he proposes a CSL dynamical equation
in which the operators $\xi^\dagger (x) \xi (x)$ are replaced by the energy
density operators $(K^{1/2} \xi^\dagger (x) ) ( K^{1/2} \xi (x) )$,
where $K^2 = - \nabla^2 + M^2$ and $\xi (x)$ is the annhilation
operator of a particle of mass $M$ at location $x$.\footnote{As
  Pearle notes, alternative energy density operators could be used.}
(Pearle's notation suppresses spin degrees of freedom.) 
This gives \cite{pearle2018dynamical} the density matrix evolution equation
\begin{equation}\label{pearleenergycsl}
{  {\partial } \over {\partial t}} \rho (t ) = - i [ H, \rho(t) ] - {
  \lambda \over {2 M_N^2 }} \int dx \int dx' \exp ( - { {
    (x-x' )^2 } \over {4 a^2 }} ) [ K^{1/2} \xi^{\dagger} (x) K^{1/2}
\xi (x) , [ K^{1/2} \xi^\dagger (x' ) K^{1/2} \xi (x' ), \rho (t) ]] \, ,
\end{equation}
where $M_N$ is the neutron mass, $\lambda$ the collapse rate and $a$
the collapse range.  The vectors $x,x'$ and integrals $\int dx , \int dx'$
are three-dimensional; similarly for momenta below.  

As we are considering photon-only collapse here, we need to reconsider
the motivations for various possible types of collapse postulate.

\subsection{Collapse based on approximate energy density}

We propose photon-only collapse models based on approximate photon energy
density, i.e. by a version of Eqn. (\ref{photonenergycsl}) in
which -- 
differently from Ref. \cite{pearle2018dynamical} -- the collapse terms involving quantities
related to $K$ and $\xi$ are defined only for photons.   
Explicitly,

\begin{equation}\label{photonenergycsl}
{  {\partial } \over {\partial t}} \rho_{AB}  (t ) = - i [ H, \rho_{AB}
(t) ] - {
  \lambda \over {2 M_N^2 }} \int dx \int dx' \exp ( - { {
    (x-x' )^2 } \over {4 a^2 }} ) [ K^{1/2} \xi^{\dagger} (x) K^{1/2}
\xi (x) ,  [ K^{1/2} \xi^\dagger (x' ) K^{1/2} \xi (x' ), \rho_{AB} (t)
]] \, ,
\end{equation}
where we write $\rho_{AB} (t)$ to emphasize that the density
operator represents the complete physical system, defined on $H_A \otimes H_B$.
Here $H_A , H_B$ represent matter and electromagnetic degrees
of freedom respectively, $K^{1/2}$ is as above with $M=0$ and $\xi (x)$ is the
photon detection or annihilation operator which, with explicit spin summation, is given by
\begin{equation}
\xi(x) = \sum_s \int_k a_{k,s} \epsilon_{k,s} \exp( i { k
  \cdot x } ) dk
\end{equation}

\subsection{Collapse based on approximate photon number} 

We argue that photon-only collapse
models based on approximate photon number density are also well motivated.
Photons are not localizable, and there is no density operator whose integral
precisely counts the number of photons in a volume.
Nonetheless, there are useful approximations to the concept \cite{mandel1966configuration}.  

First, note that in models that use non-relativistic photon-only collapse, in idealised   
environments where photon number is conserved, we can treat photons as free particles that
scatter from matter without absorption.   Treating massive particles
as qualitatively different from photons then avoids the need to discuss
further the distinction between elementary and composite particles,
which was the main initial motivation for focussing on mass-density collapse
models for massive particles.  Of course, this leaves questions about how 
virtual photons would be treated in relativistic models.
However, similar questions apply to relativistic versions
of any dynamical collapse model.   Our main aim here is to propose
alternatives to the familiar non-relativistic dynamical collapse models,
leaving discussion of relativistic versions for future work.

Second, although experimental data tend to favour conventional
mass-density collapse CSL models over particle number collapse CSL
models, the latter are not completely excluded \cite{bassi2013models},
and in any case the same conclusion does not necessarily apply
for photon-only collapse models.  Both types of photon-only collapse model have different
properties from their familiar CSL counterparts, and their empirical
implications need to be examined independently and afresh.

Third, although it seems reasonable to motivate conventional mass-density
collapse CSL  models as potentially more natural candidates
for some unified theory involving gravity, other ideas
in this direction are also worth exploring, particularly given that even consistent
(special) relativistic CSL models have not so far been constructed.
For example, gravity could plausibly be coupled to mass density
in CSL models in which the collapse operators are defined by other
quantities.   Another interesting idea, which we discuss below, is that quantum theory
and gravity are linked by graviton-only collapse models.
As in the photon case, these could involve either
particle number or energy density collapse.

In summary, photon-only number density collapse
models appear worth investigating. 
We thus want to consider the evolution equation 
\begin{equation}\label{photonnumbercsl}
{  {\partial } \over {\partial t}} \rho_{AB} (t ) = - i [ H, \rho_{AB}(t) ] - {
  \lambda } \int dx \int dx' \exp ( - { {
    (x-x' )^2 } \over {4 a^2 }} ) [ \xi^{\dagger} (x) \xi (x) , [
\xi^\dagger (x' ) \xi (x' ), \rho (t) ]] \, .
\end{equation}
Here $\lambda$ is the collapse rate, $a$ the collapse range,
$\rho_{AB}(t)$ is a density matrix for the complete physical system
(matter and photons), but the collapse terms involve only
the photon annihilation and creation operators, $\xi$ and
$\xi^\dagger$.

\section{Effects of photon collapse}

\subsection{Effects on matter}

Most lab experiments and most direct observations of matter
take place in rich and complex photonic environments.
This is also true of the retina and brain, which give
rise to our perceptions of definite outcomes of
quantum experiments and of matter apparently behaving
classically.   
A detailed analysis of the effects of photon collapse models needs, inter alia,
a detailed description of the initial environment state,
the effects on that state of scattering from the
matter system in question, the effects of subsequent
re-scatterings from it and other matter systems,
and a careful discussion of how photon number or energy density
collapses on the possible scattered states distinguish between
different matter states and so induce their effective collapse.

The simplest non-relativistic photon collapse models
conserve photon number and so do not include emission or absorption,
but more realistic models clearly should.  A simple option would be to
draw some (ultimately imprecise) distinction between photons that are generally
regarded as ``free'' in between their emission and
absorption (such as a laser pulse fired at an absorbing wall)
and virtual photons.    One could then model the dynamics
to allow varying numbers of photons and massive systems with
variable energy, and consider the effects of photon
collapse in these models.

Our aim here is to get tentative initial insight, using very rough estimates, into whether
photon-only collapse models may be consistent with present empirical
evidence and candidate solutions to the measurement problem.
To do this, we consider discrete collapse models related to
the continuous models (\ref{photonnumbercsl}) and (\ref{photonenergycsl}). 
Very roughly speaking, these have the effect of approximate
measurements of a smeared version of the number of photons of wavelengths
$\lambda \ll a$, or the total energy of such photons,
within a sphere of diameter $a$.

\subsubsection{Photon number collapse}

Define
\begin{equation}
N(a, {\bf x}) = \int \exp( - ({\bf x'}-{\bf x})^2 / a^2 ) \xi^{\dagger} ({\bf x'}) \xi
({\bf x'}) d^3 {\bf x'} \, . 
\end{equation}
Loosely speaking, this approximately represents a Gaussian weighted integral of the 
number of photons of wavelengths $\ll a$ in a volume centred at $x$.
We can define \cite{ghirardi1990markov} a discrete collapse model in which at each ${\bf x}$
discrete collapse processes takes place at random times, with the
effect
that
\begin{equation}
  \psi \rightarrow \psi^{\bf x}_n = {{ \phi^{\bf x}_n } \over { || \phi^{\bf x}_n || }} \, ,
\end{equation}
where
\begin{equation}
 \phi^{\bf x}_n = (b/ \pi)^{1/4} \exp ( - (b/2) ( N_a ({\bf x } ) - n )^2 ) \psi  \, .
\end{equation}
These collapses take place with frequency density $\mu$, so that 
that the probability of a collapse taking place for
some ${\bf x}$ in a small volume $V$ during a small time interval $t$
is $\mu V t$.

We take the total volume of space to be finite
so that the collapses take place at discrete times.
Between collapses the state $\psi$ evolves unitarily.
A finite space volume could be imposed by regularizing to a cube of
edge length $L$ with periodic boundary conditions.
Here, we simply ignore collapses outside a laboratory
or given region of interest.  Since such collapses can only increase
the effective collapse rate of systems inside the region, lower bounds on collapse rates  derived by ignoring them remain
valid.  In principle, they could affect the discussion of upper bounds
on collapse rates.   However, in practice, any radiation that has left the region of
interest and remains entangled with the system of interest has in any
case decohered the system, so that collapses
affecting such radiation do not have any additional observable effect
within the region.
This decoherence is in practice generally irreversible, and special
cases where the radiation re-enters the region
and recombines with the system can be treated by enlarging
the considered region.

Roughly, then, we can take $N_a ({\bf x})$ to be the number of photons
of wavelength $\ll a$ in a sphere of radius $a$ centred at ${\bf x}$.
A collapse constrains its value to within $\sim b^{-1/2}$, and
these collapses take place within a volume $V$ every $ ( \mu V)^{-1}$s
on average.

Consider a dust grain of size $10^{-4}$m, roughly the smallest
size visible by the human eye, in a superposition of
two separated centre-of-mass position states, exposed to
sunlight, with the line between the two positions roughly
perpendicular to the direction of incident sunlight.
Although it is not precisely rigorous to speak of the number of
photons in a given volume, photodetectors do detect individual
photons of given wavelengths, and the intuition that photons
can be roughly counted is a helpful guide.
In this sense, the flux of sunlight produces $\sim 3 \times 10^6$
photons per ${\rm cm}^3 \, {\rm s}$, or $\sim 3$ photons per second in a cube of side
$10^{-4}$m. 

We crudely model the effect of sunlight scattering from the dust
grain by supposing that, inter alia, a shadow region of cross-section
$10^{-4}$m $\times 10^{-4}$m and length $1$m is created behind the
dust grain, and that in the ``shadow'' the number of photons is
diminished by ${1 \over 3}$, so that it contains $\sim 2$ photons in each
cube of side $10^{-4}$m.   In the given configuration, the
``shadows'' of the two position states of the dust grain
do not overlap.

So, if the collapses apply on the scale of cubes of size $\sim
10^{-4}$m, make the number of optical spectrum photons
definite to within $\sim 1$, and occur every $\sim 1$s,
they will distinguish the two shadows, and hence the
position states, of the dust grain within $\sim 10^{-4}$s,
which is shorter than human perception times of $\sim 10^{-2}$s.
We take the near-definite position state to be reflected
as a definite outcome in the ontology.  Thus we assume here
 some resolution of
the tails problem (for more extensive discussion of which see e.g. \cite{mcqueen2015four}).
One possible resolution is that
other small contributions to mass density in the unselected
region, which will be present in any realistic ontology, swiftly dominate the contributions from
the unselected dust grain position state \cite{kent2017quantum}.
Another possible stance is that one should not worry overly about espilonics in a
model that is at best an approximation to a more compelling underlying
theory, in which the tails problem does not arise.
Some such assumptions are standardly made to justify the claim that
standard dynamical collapse models resolve the measurement problem.
Our first goal here is to argue that environmental collapse models
should be on the table alongside standard dynamical collapse models
as potential solutions, and we see no argument that the tails problem 
is more serious for environmental collapse models, so we
leave more detailed discussion for future work. 

In summary, modulo the tails problem, the swift transition of the dust grain
into a near-definite position state is consistent with our
perceiving within $\sim 100$ms that it has an essentially definite
position.  In other words, perception and ontology are aligned.

\subsection{Effects in photon experiments}

A simple Mach-Zehnder interference experiment with
a single photon split into two beams and recombined,
with path lengths $1$m, has flight time
$\sim 3 \times 10^{-9}$s.   If we take the
paths to run through $10^{-4}$m cubes, a
collapse in a cube distinguishing between photon
number $0$ and $1$ corresponds to a detection
of a definite path, destroying the interference.
The collapse probability is the same as if the
photon stayed in a single cube for its flight time,
i.e. $P ({\rm collapse}) \sim 3 \times 10^{-9}$.
Interference statistics would need to be demonstrated
to greater than this precision to observe the
predicted anomaly.

In a recent boson sampling experiment \cite{zhong2020quantum}, $25$ photons
travel for $\sim 3$m, with a flight time $\sim 10^{-8}$s.
An anomalous outcome would thus be observed in a run
with probability $\sim 4 \times 10^{-6}$.
Given the nature of the experiment, which produces
results that are argued to be essentially impossible to reproduce
on classical computers, even if anomalies arose it would
be difficult to identify them.

Neither type of experiment thus seems likely to refute
a photon-only collapse model, with the given parameters,
in the near future.  

\subsection{Photon energy density collapse}

Although photon energy density collapse models and
photon number collapse models have different phenomenologies
in general, we can apply similar reasoning regarding their
implications in simple discrete collapse models.
A dust grain ``shadow'' that contains $\sim 1$ fewer visible
photon per $10^{-4}$m cube thereby contains $\approx h \nu$ J 
less energy on average, where $\nu$ is the frequency of
a typical sunlight photon.   These frequencies mostly range over
$4-8 \times 10^{14}$Hz.  A discrete collapse model
that makes the photon energy within a cube definite
to within $\sim h \times 8 \times 10^{14}$ J thus has
only slightly stronger (by a factor of $\lesssim 2$)
effects than the photon number model
in the situations discussed above.
The dust grain shadow model again predicts an
effective collapse of the dust grain within
$\sim 10^{-4}$s, while again predicting undetectably small
effects in Mach-Zehnder interference and boson
sampling experiments.

\subsection{Laser experiments}

Pearle \cite{pearle2018dynamical} gives detailed analyses of anomalous effects of
photon energy density collapse in possible
laser experiments.
While Pearle's discussion considers models
that involve both massive particle and photon energy density collapse,
the effects on laser beams derive solely from the latter.
Pearle concludes that, under heroic experimental assumptions,
Eqn. (\ref{pearleenergycsl}) could produce
detectable losses of photons from X-ray laser pulses
if $\lambda \gtrsim 10^{-6}$-$10^{-4} {\rm s}^{-1}$
and could produce detectable photon ``spraying'' from
high energy CW lasers, monitored for a year,
if $\lambda \gtrsim 10^{-8} {\rm s}^{-1}$.

It would be very interesting to extend these
analyses to the various models proposed above, but
we postpone this for future work. 
As far as we are aware, these experiments have not so far been
carried out, and it is not clear that -- if and when feasible --
they would provide the strongest attainable bounds.

\subsection{Constraints on collapse parameters from human perception
  times}

One can indirectly argue, as above, for lower bounds on collapse
rates by assuming that a collapse model must imply that a physical system (such as a delocalized
dust grain) should collapse within the time it takes for us to observe
it in an essentially definite and quasi-classical state.
Indeed, authoritative reviews of collapse models (e.g. \cite{bassi2013models}) often
give lower bounds derived in precisely this way.
While this appears comfortingly objective, the fact is that
our perceptions are the result of physiological processes
in the retina and brain, and that a collapse model might
possibly predict that the relevant collapses generally
occur as a result of these processes.
Indeed, in beam-splitter experiments with 
weak light pulses, or other experiments where a human eye plays
the role of photo-detector measuring a superposition of photon states \cite{albert1989proposed}, the relevant collapse
can only be caused by events after the pulse enters (or
does not enter) the retina, according to standard dynamical collapse
models.

This is essentially true also of the
photon-only collapse models discussed here.
Although in principle it is possible for a collapse
to effectively determine the path a photon follows
from beamsplitter to (or away from) eye, the
probability must be negligibly small in order
for the model to be consistent with simple
photon interferometry experiments.
If a collapse takes place within $\sim 100$ms of
the photon leaving the source, it must be a
collapse associated with differential electromagnetic fields
generated by the nervous system, brain or other parts of the
anatomy and/or with differential scattering
patterns of environmental photons from these systems.

Attempts have been made (e.g. \cite{aicardi1991dynamical,bassi2010breaking,kent2018perception}) to analyse the mesoscopic mass distribution changes
associated with visual perception in order to obtain lower bounds on
collapse rates for standard mass-dependent dynamical collapse models.
The problem is complex.  One problem is that cellular
responses to stimuli involve proteins and molecules of various
densities diffusing in cellular media, displacing other proteins
and molecules, which makes it hard to estimate the net effect
on masss density distribution.
Another is that it is not clear how far one should pursue the
effects along the processing chain (which itself is not
known with certainty) within the brain.
Beyond this, there is also a case that it might be justifiable to
consider physiological responses -- slight muscle movements,
alterations in heart rate, and so on.   These might be
unconscious and involuntary, but nonetheless predictable,
and they may involve much greater differential displacements
of mass density than retinal or neural processing does.

Similar and perhaps even more complex issues arise with
photon-only collapse models.   The intracellular release of ions from
gates certainly generates small electromagnetic fields and has
some effect on differential scattering.   Brain processing as
a whole generates measurable electromagnetic fields.
Differential physiological responses would certainly
generate differential scattering patterns.
Any responses involving more than microscopic motion
create mass shifts at the body boundary
significantly greater than those of a displaced $10^{-4}$m dust grain.
Although our shadow model does not directly apply, similarly modelling
differentially reflected radiation suggests 
that differential scattering would induce effective collapse
on timescales short compared to $\sim 100$ms.
Of course, detailed analyses of all these effects are needed.
That said, at least if it is considered legitimate to allow
for possible physiological responses, the relatively
macroscopic scale of these responses suggests that 
our model parameters are adequate to resolve the measurement
problem.

\section{Graviton collapse}

The plausibility of photon-based collapse models suggests
considering analogous models base on graviton collapse.
Intuitively, the idea that collapse is associated with
gravity is attractive, offering the hope that the
quantum measurement problem and the conceptual problems
of quantum gravity could be solved together.
Versions of this intuition have been set out by
Diosi \cite{diosi1987universal} and Penrose \cite{penrose1996gravity}.   Diosi's and Penrose's
proposals appear to have been empirically refuted by
recent experimental data \cite{donadi2020}, suggesting
that new ideas in this direction are needed \cite{donadi2020}.

Collapse models in which the collapse terms only
directly apply to gravitons seem good candidates.
Rather than proposing, as Penrose and Diosi do, that
matter undergoes collapses to prevent macroscopic
superpositions, we propose that collapses take
place in the quantum gravitational field itself.
For an appropriate collapse rate, this offers an alternative mechanism
by which superpositions
of significantly distinct space-times could
be suppressed.  Intuitively, one can imagine physics taking place in
an approximately classical background space-time
that defines a local causal structure but
also admits small quantum fluctuations.
Because the postulated collapses do not directly
apply to matter, the analyses of Ref. \cite{donadi2020},
and specifically their Eqn. (3), do not apply. 
The anomalous photon emission from matter
implied by these analyses, and excluded by the
Gran Sasso experiment \cite{donadi2020}, need not arise.

Penrose's proposal that superpositions of space-times differing by one graviton
should quickly collapse, although difficult to make precise, is an interesting
and intuitively attractive criterion for defining collapse rates.    
However, it is also interesting to consider the full range of possible
collapse rates in graviton-only collapse models. 
Experimental upper bounds on graviton collapse
rates analogous to those (actually or potentially)
attainable for photons are  not available, as 
we have no graviton interference experiments, nor
graviton lasers.  
The viability of graviton-only collapse is nonetheless partly testable,
given specific quantum gravity models, from
constraints based on human perception.
The effective (collapse-associated) decoherence rate that graviton-based
collapse implies for humans in different perception
states cannot be greater than the total (standard) decoherence
implied by tracing out gravitational degrees of freedom.
Since the differences in gravitational fields associated
with different brain perception states are very small,
this is already quite a strong constraint, if we restrict
attention to brain processing events.
Physiological responses produce more strongly differentiated
gravitational fields, but even so, whether their scale
and speed is sufficient to induce gravitational
decoherence within $\sim 100$ms is likely model-dependent.
We hope to present detailed analyses elsewhere \cite{akbraindecoherenceestimates}.

One might even further speculate that a slight anomalous energy
production in the gravitational field created by graviton collapse
could somehow be connected to dark energy.   

\section{Conclusions}

We have raised the question of whether photon- or graviton-only
collapse models might solve the quantum measurement problem
while predicting sufficiently small deviations from standard
quantum theory that they are consistent with known experiment.
Our preliminary and very rough discussions have not
found immediately obvious flaws with this idea.
Since it has some attractive features,
we put it on the table and plan to explore it
further.

We have discussed only non-relativistic models.
This does not disfavour our models compared to standard dynamical
collapse models,
which are not presently known to have fully consistent relativistic
extensions.  Indeed, relativistic
versions of photon-only collapse models might perhaps be easier to construct.   

Lower bounds on collapse rates in any collapse model are difficult
to justify precisely, since they ultimately depend on human
perceptions of definiteness.   These are produced by very complex
physical systems interacting strongly with quite complex environments,
which make precise calculations difficult.   Potentially questionable
assumptions also need to be made.  For example, one needs to decide
whether our impression that we perceive definite outcomes within
$\sim 50-100$ms is reliable, or if not, whether there is some
reliable timescale; whether the collapse ontology should support
this perception for typical humans in typical environments or
all humans in all conceivable environments; whether possible physiological
responses should be considered relevant or whether one should restrict
analysis to brain processing events.
Still, there are reasonable justifications for particular sets of
assumptions (for example, $\sim 50-100$ms is reliable, a collapse
ontology should be aligned with this at least in typical cases,
physiological responses may (or may not) be relevant), and
the questions are fundamentally interesting enough to motivate tackling
the complexities.

\section{Acknowledgements}
I gratefully acknowledge the support of a Foundational Questions
Institute  (FQXi) grant. 
This work was partially supported  by Perimeter Institute
for Theoretical Physics. Research at Perimeter Institute is supported
by the Government of Canada through Industry Canada and by the
Province of Ontario through the Ministry of Research and Innovation.

\section*{References}

\bibliographystyle{unsrtnat}
\bibliography{photonreality,qualia3}{}
\end{document}